\documentclass[3p,times,twocolumn]{elsarticle}

\usepackage{ecrc}

\volume{00}

\firstpage{1}

\journalname{Nuclear Physics B Proceedings Supplement}

\runauth{G. Altarelli}

\jid{nuphbp}

\jnltitlelogo{Nuclear Physics B Proceedings Supplement}

\usepackage{amssymb}
\usepackage{amsmath}
\usepackage{hyperref}     
\usepackage{graphicx}

\usepackage[figuresright]{rotating}

\newcommand{\BBar}{$B^0_s$-${\kern 0.18em\overline{\kern -0.18em B^0_s}}$}

\begin{document}
\begin{frontmatter}




\title{An Overview of Neutrino Mixing}


\author{G. Altarelli}
\ead{guido.altarelli@cern.ch}
\address{Dipartimento di Fisica `E.~Amaldi', Universit\`a di Roma Tre, \\INFN Sezione di Roma Tre, I-00146 Rome, Italy\\
and\\
CERN, Department of Physics, Theory Unit, CH-1211 Geneva 23, Switzerland} 

\begin{abstract}
We present a concise review of the recent important experimental developments on neutrino mixing (hints for sterile neutrinos, large $\theta_{13}$, possible non maximal  $\theta_{23}$, approaching sensitivity on $\delta_{CP}$) and their implications on models of neutrino mixing. The new data disfavour many models but the surviving ones still span a wide range going from Anarchy (no structure, no symmetry in the lepton sector) to a maximum of symmetry, as for the models based on discrete non-abelian flavour groups.
\end{abstract}


\end{frontmatter}


\section{Introduction}
\label{}

On the experimental side the main recent developments on neutrino mixing \cite{Altarelli:2004za,Mohapatra:2006gs,Grimus:2006nb,GonzalezGarcia:2007ib,Altarelli:2010fk} were the results on $\theta_{13}$ from T2K\cite{Abe:2011sj}, MINOS\cite{Adamson:2011qu}, DOUBLE CHOOZ\cite{Abe:2011fz}, RENO \cite{Ahn:2012nd} and especially Daya Bay \cite{An:2012eh}  (see Table 1, where the latest numbers presented at the summer '12 conferences were collected). The combined value of $\theta_{13}$ is by now about 10 $\sigma$ away from zero and the central value is rather large. In turn a sizable $\theta_{13}$ allows to extract an estimate of  $\theta_{23}$ from accelerator data like T2K and MINOS. There are solid indications of a deviation of $\theta_{23}$ from the maximal value, in the first octant. In addition, some tenuous hints that $\cos{\delta_{CP}}<0$ are starting to appear in the data. A summary of recent global fits to the data on oscillation parameters is presented in Table 2  \cite{Fogli:2012ua}, \cite{GonzalezGarcia:2012}, \cite{Tortola:2012te}.

\begin{table}[h!]
\begin{center}
\begin{tabular}{|c|c|c|}
  \hline
  &&\\[-3mm]
  Quantity 	& $\sin^22\theta_{13}$ & $\sin^2\theta_{13}$ \\[1mm]
  \hline
  &&\\[-2mm]
  T2K\cite{Abe:2011sj} 					& 	$0.11^{+0.11}_{-0.06}$ 				
  													& 	$0.028^{+0.019}_{-0.024}$  \\[1mm]
  MINOS\cite{Adamson:2011qu} 	&	$0.041^{+0.047}_{-0.031}$ 		
  													& 	$0.010^{+0.012}_{-0.008}$ \\[1mm]
  DC\cite{Abe:2011fz} 				&	$0.0109\pm0.039$ 				& $0.028\pm0.011$ 			\\[1mm]
  RENO\cite{Ahn:2012nd}		&	$0.113\pm0.023$				& $0.029\pm0.006$ 			\\[1mm]
  DYB\cite{An:2012eh}					&	$0.0890\pm0.0112$				& $0.023\pm0.003$ 							\\[1mm]
  \hline
  \end{tabular}
\caption{The reactor angle measurements from the recent experiments T2K\cite{Abe:2011sj}, MINOS\cite{Adamson:2011qu}, DOUBLE CHOOZ\cite{Abe:2011fz}, Daya Bay \cite{An:2012eh} and RENO \cite{Ahn:2012nd}, for the normal hierarchy case (in the inverse hierarchy case the values do not differ by much)}
\end{center}
\end{table}

\begin{table}[h]
\begin{center}
\begin{tabular}{|c|c|c|}
  \hline
  Quantity & Ref. \cite{Fogli:2012ua} & Ref. \cite{GonzalezGarcia:2012} \\
  \hline
  $\Delta m^2_{sun}~(10^{-5}~{\rm eV}^2)$ &$7.54^{+0.26}_{-0.22}$ & $7.50\pm0.185$  \\
  $\Delta m^2_{atm}~(10^{-3}~{\rm eV}^2)$ &$2.43^{+0.06}_{-0.10}$ & $2.47^{+0.069}_{-0.067}$  \\
  $\sin^2\theta_{12}$ &$0.307^{+0.018}_{-0.016}$ & $0.30\pm0.013$ \\
  $\sin^2\theta_{23}$ &$0.386^{+0.024}_{-0.021}$ &  $0.41^{+0.037}_{-0.025}$ \\
  $\sin^2\theta_{13}$ &$0.0241\pm0.025$ &$0.023\pm0.0023$  \\
  \hline
  \end{tabular}
\end{center}
\caption{Fits to neutrino oscillation data. For $\sin^2\theta_{23}$ only the absolute minimum in the first octant is shown}
\end{table}

On the possible existence of sterile neutrinos a number of hints have been recently reported. They do not make yet an evidence but certainly pose an experimental problem that needs clarification. First, there is the MiniBooNE experiment  that has recently published \cite{Mboo:2012} a combined analysis of  $\nu_e$ appearance in a $\nu_\mu$ beam together with  $\bar{\nu}_e$ appearance in a $\bar{\nu}_\mu$ beam. An excess was observed for events with (anti)neutrino energy in the range $200<E_\nu<3000$ MeV (3.8$\sigma$). The allowed regions resulting from a two-neutrino fit of the data are consistent with $\nu_\mu \rightarrow \nu_e$ and $\bar{\nu}_\mu \rightarrow \bar{\nu}_e$ oscillations in the 0.01 to 1 $eV^2$ range for $\Delta m^2$ and consistent with the allowed region reported by the LSND experiment. In both channels, but especially in the neutrino beam case, most of the excess is at low $E_\nu$, below $~500$ MeV, where the evaluation of the background is particularly difficult. Recently the ICARUS experiment at Gran Sasso has published the results of a search for electrons produced by the CERN beam \cite{Antonello:2012}. No excess over the background was observed. As a consequence most of the region allowed by LSND, MiniBooNE. KARMEN... is excluded and only a small domain around $(\Delta m^2, \sin^2 (2\theta)) = (0.5eV^2, 0.05))$ is left. Then there is the reactor anomaly: a reevaluation of the reactor flux \cite{Mention:2011} produced an apparent gap between the theoretical expectations and the data taken at small distances from  reactors ($\le$ 100 m). The discrepancy is of the same order of the quoted systematic error whose estimate, detailed in the paper, should perhaps be reconsidered. Actually, a different analysis confirmed the normalization shift \cite{Huber:2011}. Similarly the Gallium anomaly \cite{Giunti:2010} depends on the assumed cross-section which could be questioned. Cosmological data allow, actually even favour,  the existence of one sterile neutrino, while the most stringent bounds arising form nucleosynthesis disfavour two or more sterile neutrinos (assuming that they are thermalized) \cite{Giusarma:2011,Joudaki:2012}. Over all, only a small leakage from active to sterile neutrinos is allowed by present neutrino oscillation data (see, for example, refs. \cite{Giunti:2011a,Giunti:2011b,Archidiacono:2012} and references therein). If all the indications listed above were confirmed (it looks unlikely) then adding one or more sterile neutrinos would probably not be enough to satisfactorily describe all the data. There is in fact a strong tension between appearance (LSND, MiniBooNE, ICARUS) and disappearance (reactors, Ga anomaly) data. Thus the situation is at present very confuse but the experimental effort should be continued because establishing the existence of sterile neutrinos would be a great discovery. In fact a sterile neutrino is an exotic particle not predicted by the most popular models of new physics. In the following we restrict our discussion to 3-neutrino models.

The rather large measured value of $\theta_{13}$, close to the old CHOOZ bound and to the Cabibbo angle, and the indication that $\theta_{23}$ is not maximal both go in the direction of models based on Anarchy \cite{Hall:1999sn,deGouvea:2003xe}, i.e. the ansatz that perhaps no symmetry is needed in the leptonic sector, only chance (this possibility has been recently reiterated, for example, in Ref. \cite{deGouvea:2012ac}). Anarchy can be formulated in a $SU(5) \otimes U(1)_{FN}$ context by taking different Froggatt-Nielsen \cite{Froggatt:1978nt} charges only for the $SU(5)$ tenplets (for example $10\sim(a,b,0)$, where $a > b > 0$ is the charge of the first generation, b of the second, zero of the third) while no charge differences appear in the $\bar 5$ (e. g. $\bar 5\sim (0,0,0)$).  The observed fact that the up-quark mass hierarchies are more pronounced than for down-quark and charged leptons is in agreement with this assignment. The $SU(5)$ generators act
ÔverticallyÕ inside one generation, whereas the $U(1)_{FN}$ charges are different ÔhorizontallyÕ from one
generation to the other. If, for a given interaction vertex, the $U(1)_{FN}$  charges do not add to zero, the
vertex is forbidden in the symmetric limit. However, the $U(1)_{FN}$ symmetry (that one can assume to be a gauge symmetry) is spontaneously broken by
the VEVs $v_f$ of a number of ÔflavonÕ fields with non-vanishing charge and GUT-scale masses. Then a forbidden coupling
is rescued but is suppressed by powers of the small parameters $\lambda = v_f/M$, with $M$ a large mass, with the exponents larger for larger charge mismatch. Thus the charges fix the powers of $\lambda$, hence the degree of suppression of all elements of mass matrices, while arbitrary coefficients $k_{ij}$ of order 1 in each entry of mass matrices are left unspecified (so that the number of parameters exceeds the number of observable quantities). A random selection of these $k_{ij}$ parameters leads to distributions of resulting values for the measurable quantities. For anarchy (A) the mass matrices in the leptonic sector are totally random, while in the presence of unequal charges different entries carry different powers of the order parameter and thus some hierarchies are enforced. The embedding of Anarchy in the $SU(5) \otimes U(1)_{FN}$ context allows to implement a parallel  treatment of quarks and leptons. Within this framework there are many variants of these models: fermion charges can all be nonnegative
with only negatively charged flavons, or there can be fermion charges of different signs
with either flavons of both charges or only flavons of one charge.
In models with no see-saw, the $\bar 5$ charges completely fix the hierarchies (or anarchy, if the case) in the neutrino mass matrix.  If Right-Handed (RH) neutrinos are added, they transform as $SU(5)$ singlets and can in principle carry $U(1)_{FN}$ charges, which also must be all equal in the anarchy case. With RH neutrinos the see-saw mechanism can take place and the resulting phenomenology is modified. In Ref.\cite{AFMM:2012}, given the new experimental results, we have made a reappraisal of Anarchy and its variants within the (SUSY) $SU(5)\times U(1)_{\rm FN}$ GUT framework. Based on the most recent data we argue  that the Anarchy ansatz is probably oversimplified and, in any case, not compelling. In fact, suitable differences of $U(1)_{FN}$ charges, if also introduced within pentaplets and singlets, lead to distributions that are in better agreement with the data with the same number of random parameters as for Anarchy. The hierarchy of quark masses and mixing and of charged lepton masses in all cases impose a hierarchy defining parameter of the order of $\lambda_C=\sin{\theta_C}$, with $\theta_C$ being the Cabibbo angle. The weak points of Anarchy ($A$) are that with this ansatz all mixing angles should be of the same order, so that the relative smallness of $\theta_{13}\sim o(\lambda_C)$ is not automatic. Similarly the smallness of $r=\Delta m^2_{solar}/\Delta m^2_{atm}$ is not easily reproduced: with no See-Saw $r$ is of $o(1)$, while in the See-Saw version of Anarchy the problem is only partially alleviated by the spreading of the neutrino mass distributions that follows from the product of three matrix factors in the See-Saw formula. An advantage is already obtained if Anarchy is only restricted to the 23 sector of leptons as in the $A_{\mu \tau}$ model (in the notation of Ref.\cite{AFMM:2012}). In this case, with or without See-Saw,  $\theta_{13}$ is naturally suppressed and, with a single fine tuning one gets both $\theta_{12}$ large and $r$ small (this model was also recently rediscussed in Ref. \cite{Buchmuller:2011tm}). Actually in Ref.\cite{AFMM:2012} we have shown that, in the no See-Saw case, a very good performance is observed in a new model, the $H$ model, where Anarchy is also relaxed in the 23 sector. In the $H$ model, by taking a relatively large order parameter, one can reproduce the correct size for all mixing angles and mass ratios. Alternatively, in the See-Saw case, we have shown that the freedom of adopting RH neutrino charges of  both signs, as in the $PA_{\mu \tau}$ model, can be used to obtain a completely natural model where all small quantities are suppressed by the appropriate power of $\lambda$. In this model a lopsided Dirac mass matrix is combined with a generic Majorana matrix to produce a neutrino mass matrix where the 23 subdeterminant is suppressed and thus $r$ is naturally small with unsuppressed $\theta_{23}$. In addition  $\theta_{12}$ is large while $\theta_{13}$ is suppressed. We stress again that the number of random parameters is the same in all these models: one coefficient of $o((1)$ for every matrix element. Moreover, with an appropriate choice of charges, it is not only possible to reproduce the charged fermion hierarchies and the quark mixing, but also the order of magnitude of all small observed parameters can be naturally guaranteed.  In conclusion, we agree that models based on chance are still perfectly viable, but we consider Anarchy a particularly simple choice perhaps oversimplified and certainly not compelling and we have argued in favour of less chaotic solutions.

Anarchy and its variants, all sharing the dominance of randomness in the lepton sector, are to be confronted with models with a richer dynamical structure, in particular those based on discrete flavour groups (for a review, see, for example, Ref.~\cite{Altarelli:2010gt}). After the measurement of a relatively large value for $\theta_{13}$ there has been an intense work to interpret these new results along different approaches and ideas. An updated list of references can be found, for example, in the recent papers Refs.~\cite{Varzielas:2012ss,King:2012}. Among the models with a non trivial dynamical structure those based on discrete flavour groups were motivated by the fact that the data suggest some special mixing patterns as good first approximations like Tri-Bimaximal (TB) or Golden Ratio (GR) or Bi-Maximal (BM) mixing, for example. The corresponding mixing matrices all have $\sin^2{\theta_{23}}=1/2$, $\sin^2{\theta_{13}}=0$, values that are good approximations to the data (although less so since the most recent data), and differ by the value of the solar angle $\sin^2{\theta_{12}}$. The observed $\sin^2{\theta_{12}}$, the best measured mixing angle,  is very close, from below, to the so called Tri-Bimaximal (TB) value \cite{Harrison:2002er,Harrison:2002kp,Xing:2002sw,Harrison:2002et,Harrison:2003aw} of $\sin^2{\theta_{12}}=1/3$. Alternatively, it is also very close, from above, to the Golden Ratio (GR) value \cite{Kajiyama:2007gx,Everett:2008et,Ding:2011cm,Feruglio:2011qq} $\sin^2{\theta_{12}}=\frac{1}{\sqrt{5}\,\phi} = \frac{2}{5+\sqrt{5}}\sim 0.276$, where $\phi= (1+\sqrt{5})/2$ is the GR (for a different connection to the GR, see Refs.~\cite{Rodejohann:2008ir,Adulpravitchai:2009bg}). On a different perspective, one has also considered models with Bi-Maximal (BM) mixing, where $\sin^2{\theta_{12}}=1/2$, i.e. also maximal, as the neutrino mixing matrix before diagonalization of charged leptons.  One can think of models where a suitable symmetry enforces BM mixing in the neutrino sector at leading order (LO) and the necessary, rather large, corrective terms to $\theta_{12}$ arise from the diagonalization of the charged lepton mass matrices (a list of references can be found in Ref.~\cite{Altarelli:2010gt}). Thus, if one or the other of these coincidences is taken seriously, models where TB or GR or BM mixing is naturally predicted provide a good first approximation (but these hints cannot all be relevant and it is well possible that none is).
The corresponding mixing matrices have the form of rotations with fixed special angles. Thus one is naturally led to discrete flavour groups. Models based on discrete flavour symmetries, like $A_4$ or $S_4$, have been proposed in this context and widely studied. In these models the starting Leading Order (LO) approximation is completely fixed (no chance), but the Next to LO (NLO) corrections still introduce a number of undetermined parameters, although in general much less numerous than for $U(1)_{FN}$ models. These models are therefore more predictive and typically, in each model, one obtains relations among the departures of the three mixing angles from the LO patterns, restrictions on the CP violation phase $\delta_{CP}$, mass sum rules among the neutrino mass eigenvalues, definite ranges for the neutrinoless beta decay effective Majorana mass and so on. 

In the following we will mainly refer to TB or BM mixing which are the most studied first approximations to the data. A simplest symmetry that, at LO, leads to TB is $A_4$ while BM can be obtained from $S_4$. Starting with the ground breaking paper in Ref. \cite{Ma:2001},
$A_4$ models  have been widely studied  (for a recent review and a list of references, see Ref.~\cite{Altarelli:2012}). At LO the typical $A_4$ model (like, for example, the one discussed in in Ref. \cite{Altarelli:2006ty}) leads to exact TB mixing.  The LO approximation is then corrected by non-leading effects. Given the set of flavour symmetries and having specified the field content, the non-leading corrections to TB mixing, arising from higher dimensional effective operators, can be evaluated in a well-defined expansion. In the absence of specific dynamical tricks, in a generic model all three mixing angles receive corrections of the same order of magnitude. Since the experimentally allowed departures of
 $\theta_{12}$ from the TB value, $\sin^2{\theta_{12}}=1/3$, are small, numerically not larger than $\mathcal{O}(\lambda_C^2)$ where $\lambda_C=\sin\theta_C$, it follows that both $\theta_{13}$ and the deviation of $\theta_{23}$ from the maximal value are also expected to be typically of the same general size. The same qualitative conclusion also applies to $A_5$ models with GR mixing. This generic prediction of a small $\theta_{13}$, numerically of  $\mathcal{O}(\lambda_C^2)$, can now be confronted with the present precise experimental value. The central value $\sin{\theta_{13}} \sim 0.15$, from Table 2, is between $\mathcal{O}(\lambda_C^2) \sim \mathcal{O}(0.05)$ and $\mathcal{O}(\lambda_C) \sim \mathcal{O}(0.23)$. Since $\lambda_C$ is not that small, this gap is not too large and one can argue that models based on TB (or GR) mixing are still viable with preference for the lower side of the experimental range.

Of course, one can introduce some additional theoretical input to improve the value of $\theta_{13}$. In the case of $A_4$, one particularly interesting example is provided by the  Lin model \cite{Lin:2009bw} (see also Ref.~\cite{Varzielas:2010mp,Rodejohann:2012}), formulated before the T2K, MINOS, DOUBLE CHOOZ, Daya Bay and RENO results.  In the Lin model the $A_4$ symmetry breaking is arranged, by suitable additional $Z_n$ parities, in a way that the corrections to the charged lepton and the neutrino sectors are kept separated not only at LO but also at next-to-leading order (NLO). As a consequence, in a natural way the contribution to neutrino mixing from the diagonalization of the charged leptons can be of $\mathcal{O}(\lambda_C^2)$, while those in the neutrino sector of $\mathcal{O}(\lambda_C)$. In addition, in the Lin model these large corrections do not affect $\theta_{12}$ and satisfy the relation $\sin^2{\theta_{23}} =1/2 +1/\sqrt{2}\cos{\delta_{CP}} |\sin{\theta_{13}}|$, with $\delta_{CP}$ being the CKM-like CP violating phase of the lepton sector. Thus, in the Lin model the NLO corrections to the solar angle $\theta_{12}$ and to the reactor angle $\theta_{13}$ are not necessarily related. Note that, for $\theta_{23}$ in the first octant, the sign of  $\cos{\delta_{CP}}$ must be negative.

Alternatively, one can think of models where, because of a suitable symmetry,  BM mixing holds in the neutrino sector at LO and the corrective terms for $\theta_{12}$, which in this case are required to be large, arise from the diagonalization of charged lepton masses. These terms from the charged lepton sector, numerically of order $\mathcal{O}(\lambda_C)$, would then generically also affect $\theta_{13}$ and the resulting angle could well be compatible with the measured value. An explicit model of this type based on the group $S_4$ has been developed in Ref.~\cite{Altarelli:2009gn} (see also Refs.~\cite{Toorop:2010yh,Patel:2010hr,Meloni:2011fx}). An important feature of this model is that only $\theta_{12}$ and $\theta_{13}$ are corrected by terms of $\mathcal{O}(\lambda_C)$ while $\theta_{23}$ is unchanged at this order. This model is compatible with present data and clearly prefers the upper range of the present experimental result for $\theta_{13}$. 

In Ref. \cite{AFMS:2012} we discuss three possible classes of models: 1) typical $A_4$ models where $\theta_{13}$ is generically expected to be small, of the order of the observed departures of $\theta_{12}$ from the TB value, and thus with preference for the lower end of the allowed experimental range. 2) special $A_4$ models, like the Lin model, where $\theta_{13}$ is made independent of the deviation of $\sin^2{\theta_{12}}$ from the TB value $1/3$ and can be as large as the upper end of the allowed experimental range. In the same paper we discuss a general characterization of these special $A_4$ models where the dominant corrections to TB mixing do not arise from the charged lepton sector but from the neutrino sector. 3) Models where BM mixing holds in the neutrino sector and large corrections to $\theta_{12}$ and $\theta_{13}$ arise from the diagonalization of charged leptons. The value of $\theta_{13}$ is naturally close to the present experimental range. In each of these possible models the dominant corrections to the LO mixing pattern involve a number of parameters of the same order of magnitude,  $\xi$. We discuss the success rate corresponding to the optimal value of $\xi$ for each model, obtained by scanning the parameter space according to a similar procedure for all three cases. We argue that, while the absolute values of the success rates depend on the scanning assumptions, their relative values in the three classes of models, provide a reliable criterium for comparison. We find that, for reproducing the mixing angles, the Lin type models have the best performance, as expected, followed by the typical $A_4$  models while the BM mixing models lead to an inferior score, as they can well reproduce the size of $\theta_{13}$ but most often fail to reproduce the correct value of  $\theta_{12}$. We also discuss the conditions for $\cos{\delta_{CP}}$ peaking around -1.

In Ref. \cite{AFMS:2012} we have also discussed the implications for lepton flavour-violating (LFV) processes of the above three classes of possibilities, assuming a supersymmetric context, with or without See-Saw. The  present bounds on LFV reactions pose severe constraints on the parameter space of the models (for a recent general analysis on model-independent flavour violating effects in the context of flavour models, see Ref.~\cite{Calibbi:2012at}). In particular, we refer to the recent improved MEG result \cite{Adam:2011ch} on the $\mu \rightarrow e \gamma$ branching ratio, $Br(\mu \rightarrow e \gamma) \lesssim 2.4\times10^{-12}$ at $95\%$ C.L. and to other similar processes like $\tau \rightarrow (e~\rm{or}~ \mu)  \gamma$. One expects that lepton flavour-violating processes may also have a large discriminating power in assessing the relative merits of the different models. In Ref. \cite{AFMS:2012} we have studied this issue by adopting the simple CMSSM framework. While this very constrained version of supersymmetry is rather marginal after the results of the LHC searches, more so given that the Higgs mass is around $m_H=125$ GeV, we still believe it can be used for indicative purposes as in this case. We find that  the most constrained versions are the models with BM mixing at LO where relatively large corrections directly appear in the off-diagonal terms of the charged lepton mass matrix. The $A_4$ models turn out to be the best suited to satisfy the experimental bounds, as the non-diagonal charged lepton matrix elements needed to reproduce the mixing angles are quite smaller. An intermediate score is achieved by the models of the Lin type, where the main corrections to the mixing angles arise from the neutrino sector and the non-diagonal charged lepton matrix elements are smaller. Overall, among the discrete flavour group models that we have studied, the $A_4$ models emerge well from the analysis of Ref. \cite{AFMS:2012}and in particular those of the Lin type lead to a natural description of the data.  As for the regions of the CMSSM parameter space that are indicated by our analysis, the preference is for small $\tan{\beta}$ and large SUSY masses (at least one out of $m_0$ and $m_{1/2}$ must be above 1 TeV).  Given that SUSY has not been found at the LHC the preference for  heavy SUSY masses does not pose a problem to these flavour models, except that, as a consequence, it appears impossible, at least within the CMSSM rigid framework, to satisfy the MEG bound and simultaneously to reproduce the muon $g-2$ \cite{Bennett:2006fi} discrepancy \cite{Hagiwara:2006jt,Passera:2008jk,Passera:2008hj}.



\section*{Acknowledgement}
I would like to thank the Organizers of the Capri Workshop, in particular Giulia Ricciardi, for their kind invitation.
I recognize that this work has been partly supported by the Italian Ministero dellÕUniversit`a e della Ricerca Scientifica, under the COFIN program (PRIN 2008), by the European Commission, under the networks ÒLHCPHENONETÓ and ÒInvisiblesÓ.






\end{document}